\documentclass{aa}
\usepackage{graphicx}
\usepackage{xcolor}
\usepackage{txfonts}
\def\gsim{\,\lower4pt\hbox{${\buildrel\displaystyle >\over\sim}$}\,}
\def\lsim{\,\lower4pt\hbox{${\buildrel\displaystyle <\over\sim}$}\,}

\newcommand\rs[1]{_\mathrm{#1}}

\usepackage{graphicx}
\usepackage{xcolor}
\usepackage{txfonts}
\begin{document}
   \title{Laboratory evidence for asymmetric accretion structure upon slanted matter impact in young stars}

   \author{K. Burdonov\inst{1}, G. Revet\inst{1}, R. Bonito\inst{2}, C. Argiroffi\inst{2,3}, J. Béard\inst{4}, S. Bolanõs\inst{1}, M. Cerchez\inst{5}, S.N. Chen\inst{6}, A. Ciardi\inst{7}, G. Espinosa\inst{8}, E. Fillipov\inst{9,10},  S. Pikuz\inst{9,11}, R. Rodriguez\inst{8}, M. Šmíd\inst{12}, M. Starodubtsev\inst{10}, O. Willi\inst{5}, S. Orlando\inst{2} and J. Fuchs\inst{1}
          }

   \institute{LULI - CNRS, CEA, UPMC Univ Paris 06 : Sorbonne Université, Ecole Polytechnique, Institut Polytechnique de Paris - F-91128 Palaiseau cedex, France
   \and
   INAF, Osservatorio Astronomico di Palermo, Palermo, Italy
   \and
   Department of Physics and Chemistry, University of Palermo, 90133, Palermo, Italy
   \and
   LNCMI, UPR 3228, CNRS-UGA-UPS-INSA, 31400, Toulouse, France
   \and
   Heinrich-Heine Universität Düsseldorf, 40225, Düsseldorf, Germany
   \and
   ELI-NP, ”Horia Hulubei” National Institute for Physics and Nuclear Engineering, 30 Reactorului Street, RO-077125, Bucharest-Magurele, Romania
   \and
   Sorbonne Université, Observatoire de Paris, PSL Research University, LERMA, CNRS UMR 8112, F-75005, Paris, France
   \and
   Department of Physics, IUNAT, Universidad de Las Palmas de Gran Canaria, 35001, Las Palmas, Spain
   \and
   JIHT, Russian Academy of Sciences, 125412, Moscow, Russia
   \and
   IAP, Russian Academy of Sciences, 603950, Nizhny Novgorod,  Russia
   \and
   NRNU MEPhI, 115409, Moscow, Russia
   \and
   Helmholtz-Zentrum Dresden-Rossendorf, 01328, Dresden, Germany
   }
    
   \date{}

% \abstract{}{}{}{}{} 
% 5 {} token are mandatory
 
  \abstract
   {}
   {Investigating in the laboratory the process of matter accretion onto forming stars through scaled experiments is important in order to better understand star and planetary systems formation and evolution. Such experiments can indeed complement observations by providing access to the processes with spatial and temporal resolution. A first step has been made in \cite{2017SciA....3E0982R} in allowing such investigations. It revealed the existence of a two components stream: a hot shell surrounding a cooler inner stream. The shell was formed by matter laterally ejected upon impact and refocused by the local magnetic field. That laboratory investigation was limited to normal incidence impacts. However, in young stellar objects, complex structure of magnetic fields causes variability of the incident angles of the accretion columns. This led us to undertake an investigation, using laboratory plasmas, of the consequence of having a slanted accretion impacting a young star.}
   {Here we use high power laser interactions and strong magnetic field generation in the laboratory, complemented by numerical simulations, to study the asymmetry induced upon accretion structures when columns of matter impact the surface of young stars with an oblique angle.}
   {Compared to the scenario where matter accretes normal to the star surface, we observe strongly asymmetric plasma structure, strong lateral ejecta of matter, poor confinement of the accreted material and reduced heating compared to the normal incidence case. Thus, slanted accretion is a configuration that seems to be capable of inducing perturbations of the chromosphere and hence possibly influence the level of activity of the corona.}
   {}

   \keywords{accretion, accretion disks --
             instabilities --
             magnetohydrodynamics (MHD) --
             shock waves --
             stars: pre-main sequence --
             X-rays: stars}

\titlerunning{Laboratory evidence for asymmetric accretion structure upon slanted matter impact in young stars}
\authorrunning{K. Burdonov et~al.}

\maketitle
%
%________________________________________________________________

\section{Introduction}
The dynamics of matter accretion is a process of high interest, because of its dominant role in a wide range of astrophysical objects including the evolution of young stars and the formation of planetary systems (e.g. \citealt{doi:10.1146/annurev-astro-081915-023347, Scaringie1500686, Caratti, 2017A&A...607A..14A}). In the case of low-mass young stars, observations have provided evidence that the process of matter accretion impacts the evolution of the surrounding stellar atmosphere and, in particular, the level of coronal activity (\citealt{nss95, fdm03, sab04a, pkf05, jca06, gwj07, def09, bcd10, 2011MNRAS.415.3380O, dbc12, 2019A&A...624A..50C}). Current models also suggest possible perturbations of the stellar atmosphere by accreting streams, however up to now there were quite few studies related to such phenomena (\citealt{2010A&A...510A..71O, refId0orl, 2013A&A...557A..69M}). In the solar case, perturbations of the chromosphere were observed during the impacts of cold and dense fragments of the chromosphere previously ejected in the interplanetary medium by a solar eruption associated to a flare (\citealt{2013Sci...341..251R, 2014ApJ...797L...5R}). Perturbations of the chromosphere and corona of young stars were modeled in \cite{2010A&A...510A..71O, refId0orl}. However, the levels of perturbations were not studied in detail because the models are idealized and assume only normal impacts. A first attempt to model oblique impacts was made by \cite{2014ApJ...797L...5R} who described impacts of dense fragments onto the solar surface at an angle of 15$^{o}$ with respect to normal impacts. These studies have suggested that an oblique impact may produce much stronger perturbations of the chromosphere, possible additional heating mechanism of the overlying corona and, therefore, possible influence on the level of activity of the corona. 

In our previous work (\citealt{2017SciA....3E0982R}), we modeled normal impact of an incoming collimated plasma stream onto the surface of the star, in the presence of a magnetic field co-aligned with the stream. The plasma in the experiment was shown to be scalable to acretion events on a young star (\citealt{Ryutov_2000}). We verified in particular that our set-up is representative of a high plasma $\beta$ ($\geq 1$) Classical T Tauri Stars (CTTSs) accretion case (\citealt{REVET2019100711}). We experimentally demonstrated the formation of a shell of dense (and optically thick when scaled to astrophysical conditions) plasma which envelopes the core post-shock region and absorbs the X-rays arising from the central core. Such enveloping leads to the decreasing of X-ray flux from the star which is consistent with discrepancies between astrophysical observations and numerical predictions in the X-ray domain (\citealt{Bonito_2014}). 

Here we study this astrophysical phenomenon in a scaled laboratory experiment using laser-produced plasma streams impacting onto a tilted obstacle. These streams thus represent accretion material impacting onto a star in a slanted configuration. The formation of such streams takes place with the help of large-scale quasi-static B-field perpendicular to the surface of the target irradiated by the laser (\citealt{Albertazzi325, PhysRevLett.119.255002}). In this work we explore the oblique incidence of incoming laser-created plasma stream to the secondary obstacle target mimicking the surface of the star.

The paper is organized as follows. In Sect. \ref{sec:lab.exp} we describe the setup and approach of the laboratory experiment; in Sect. \ref{sec:lab.result} we discuss the results of the experiment and the results of the simulations and synthesis of X-ray emission; in Sect. \ref{sec:mod.result} we describe the Magnetohydrodynamics (MHD) model of accretion impacts in young stars, the numerical setup, the synthesis of the X-ray emission from the MHD model and the spectral analysis; and in Sect. \ref{sec:conc} we discuss the results and draw our conclusions.

\section{Laboratory experiment approach}
\label{sec:lab.exp}
\begin{figure}[htp]
    \centering
    \includegraphics[width=6 cm]{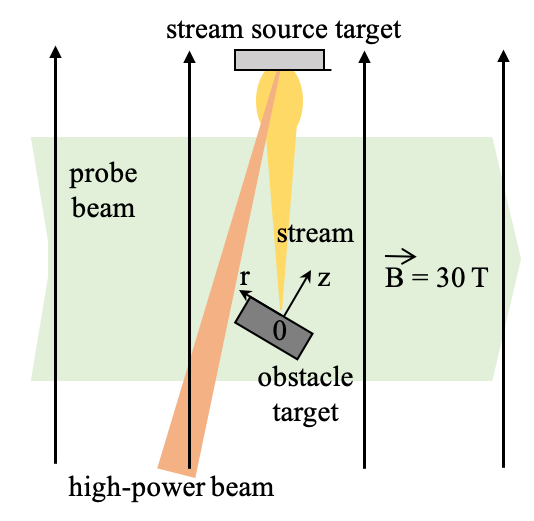}
    \caption{Schematic view of the experimental set-up.}
    \label{fig:setup}
\end{figure}

The experiments were conducted at the ELFIE laser facility (\cite{Zou_2008}), the set-up is shown in Fig. \ref{fig:setup}. An optical laser pulse with energy up to 50 J, 0.6 ns full width half maximum duration at the wavelength 1057 nm was focused onto the surface of a Teflon (CF$\rs{2}$) target using a lens with 2.2 m focal length and a random phase plate, providing a 0.5 mm focal spot diameter beam with intensity up to $4 \times 10^{13}$ W/cm$^{2}$. After irradiating the target by the laser, a hot long plasma stream was collimated by applying a large-scale quasi-static homogeneous B-field, aligned with the main expansion axis of the plasma, with a strength of up to 30 T, which was oriented normally to the target surface (\citealt{HIGGINSON201748}), as shown in Fig. \ref{fig:setup}.

This plasma stream hits the secondary obstacle PET ((C$\rs{10}$H$\rs{8}$O$\rs{4}$)$\rs{n}$) target, which was positioned at a distance of 18 mm from the first target. The incoming plasma stream had a constant diameter of ~1.0 mm at the location of the obstacle, and over more than 100 ns it impacts the obstacle with a plasma electron density ($n_{\rm e} \approx 2\times 10^{18}$ cm$^{-3}$); the measured plasma electron temperature of the incoming stream was ~10 eV or 0.1 MK. The angle at which the incoming plasma stream impacts the secondary target was set, with respect to the obstacle target normal, to 0$^{\circ}$, 15$^{\circ}$, 30$^{\circ}$ and 45$^{\circ}$ for different shots.

To investigate fully the plasma plume originated from the obstacle under the incoming stream impact, we used several complementary diagnostics, allowing us to measure plasma parameters in all principal regions of the structure.

Optical interferometry was used to investigate features of plasma propagation at distances above 0.5 mm from the obstacle target, where the plasma is transparent for the optical probe beam, providing two-dimensional side-view distributions of the plasma electron density.

The streaked optical pyrometery (SOP) diagnostic was used to record the visible light (400-600 nm) emitted, in the optically thin regime, by the plasma close to the obstacle target, i.e. in regions of higher density that cannot be accessed by the interferometry diagnostic, since the latter is obscured close to the target due to refraction of the optical probe beam induced by strong density gradients. It can hence give information on the density structure close to the obstacle surface following the impact.

A variable spacing grating (VSG) spectrometer (\citealt{Kita:83}) and a focusing spectrograph with spatial resolution (FSSR) (\citealt{Faenov_1994}) were utilized to record the emissivity of the plasma in the soft X-ray domain: broadband (0.4-2 keV) and narrow (0.8-0.95 keV) spectrum ranges respectively. As the SOP, it also resolves the plasma emission along the Z axis, but the signal is integrated along the radial (r) axis, as well as integrated in time. However, the emission is spectrally resolved in the range of sensitivity. It can hence inform on the temperature of the plasma following impact.

\section{Experimental results}
\label{sec:lab.result}
\subsection{Optical interferometry}
The experiment reveals, for slanted impact, a strong asymmetry of the plasma structure following impact, with strong lateral leakage of matter away from the impact point.

Such evolution of the plasma during the impact of the plasma flow on the obstacle was measured by means of a Mach-Zehnder based interferometer from the time ($t = 0$ ns) when the plasma stream reaches the secondary target, and up to $t = 118$ ns, which is our limit of observation corresponding to 1680 seconds (half an hour) in the astrophysical case. This number has been calculated using the laboratory-astrophysical scaling detailed in (\citealt{2017SciA....3E0982R}) which uses the same laboratory platform as used for the present experiment.
The probe beam used for the interferometer was a compressed beam with 100 mJ energy and 350 fs duration at the fundamental wavelength 1057 nm, combined collinearly with its second-harmonic pulse having 30 mJ energy and the same duration but at a wavelength of 528.5 nm. A Glan prism was used to polarizationaly split both beams into two replicas (S and P), delay the S with respect to the P, and then recombine them collinearly (see \citealt{HIGGINSON201748}). Thus before probing the plasma we had four temporally separated beams delayed relative to each other by 12 ns, which provided four interferograms in one shot.

\begin{figure*}[htp]
    \centering
    \includegraphics[width=15 cm]{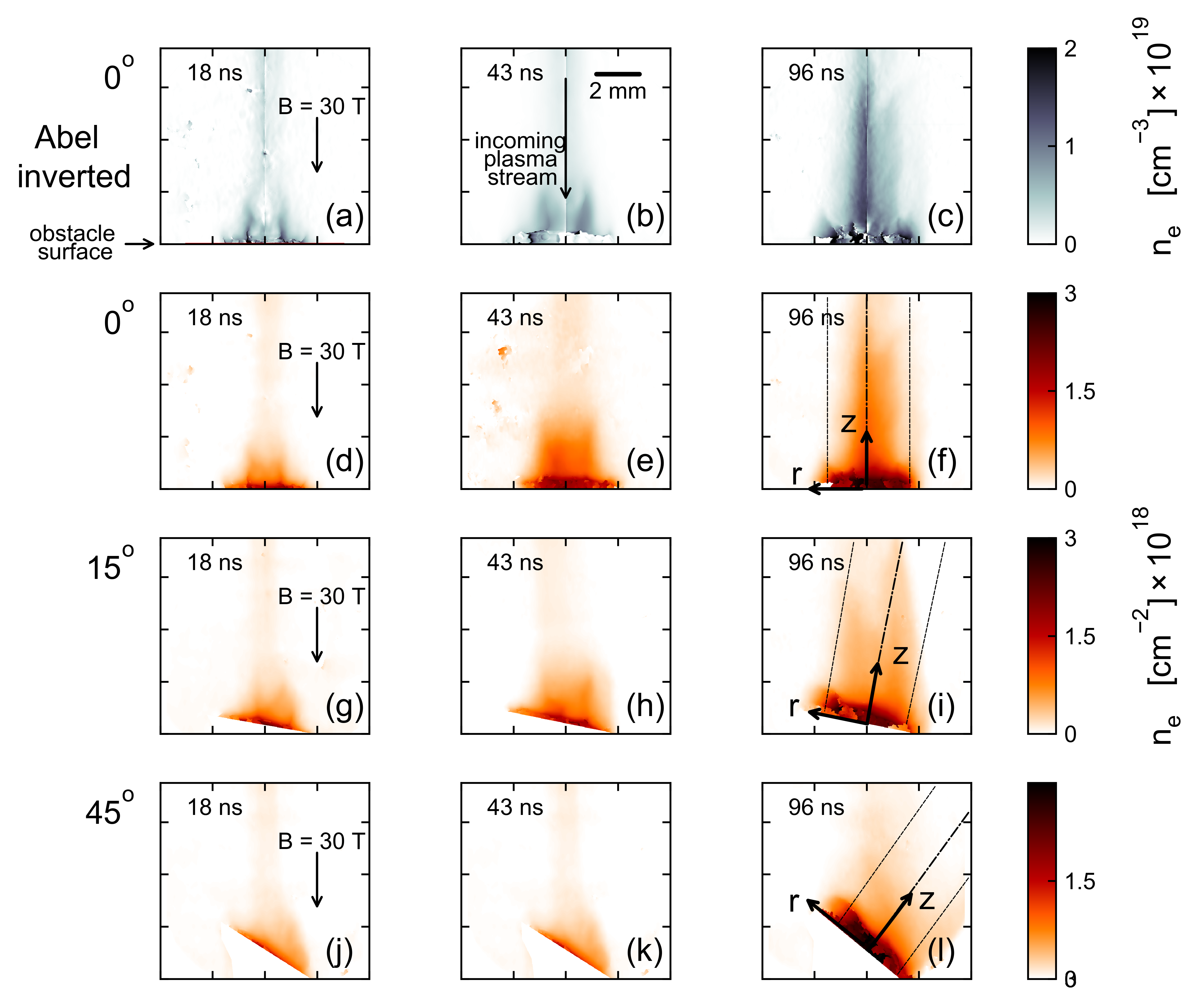}
    \caption{Evolution of plasma electron density in time (18 ns, 43 ns and 96 ns): (a)-(c) – Abel inverted volumetric density profiles for the normal incoming stream impact; (d)-(f) – two-dimensional, line-integrated, side-view distributions of the plasma electron density for normal incoming stream, corresponding to the same images as shown in panels (a)-(c); (g)-(i) – same for 15$^{\circ}$ oblique stream incidence, (j)-(l) – same for 45$^{\circ}$ oblique stream incidence.}
    \label{fig:interferometry 30T}
\end{figure*}

Plasma density snapshots for three consecutive points in time (18 ns, 43 ns and 96 ns) following the interaction of the incoming stream with the obstacle target are presented in Fig. \ref{fig:interferometry 30T}.

The initial moment ($t = 0$) and zero point ($x = 0$) matches to the time and place of the stream tip arrival to the surface of the obstacle. Figure \ref{fig:interferometry 30T} (a)-(c) shows the volumetric plasma electron density profiles retrieved by Abel inversion from the experimental phase maps obtained using time-resolved optical interferometry for the normal impact case. The corresponding two-dimensional distributions of the line-of-sight integrated density of the plasma are presented in Fig. \ref{fig:interferometry 30T} (d)-(f). Two dimensional distributions for oblique incidence angles 15$^{\circ}$ and 45$^{\circ}$ are shown in Fig. \ref{fig:interferometry 30T} (g)-(i) and Fig. \ref{fig:interferometry 30T} (j)-(l) correspondingly.

For the normal impact case, consistently with previous results (\cite{2017SciA....3E0982R}), we see the progressive formation of a symmetrical plasma structure, i.e. a characteristic shocked core within the stream and surrounded by a dense envelope, surrounding the incoming stream. This structure was presented in details in our previous work (\cite{2017SciA....3E0982R}). And for the slanted impact case, we clearly observe the development of an asymmetric plasma density distribution with a denser plasma structure on the side of the reduced angle of impact. As can be seen also in Fig. \ref{fig:interferometry 30T}, the post-shocked plasma flow has a tendency to follow the magnetic field lines.
Thus, larger obstacle tilt angles demonstrate less confined and more asymmetric plasma distributions.

\begin{figure}[htp]
    \centering
    \includegraphics[width=9 cm]{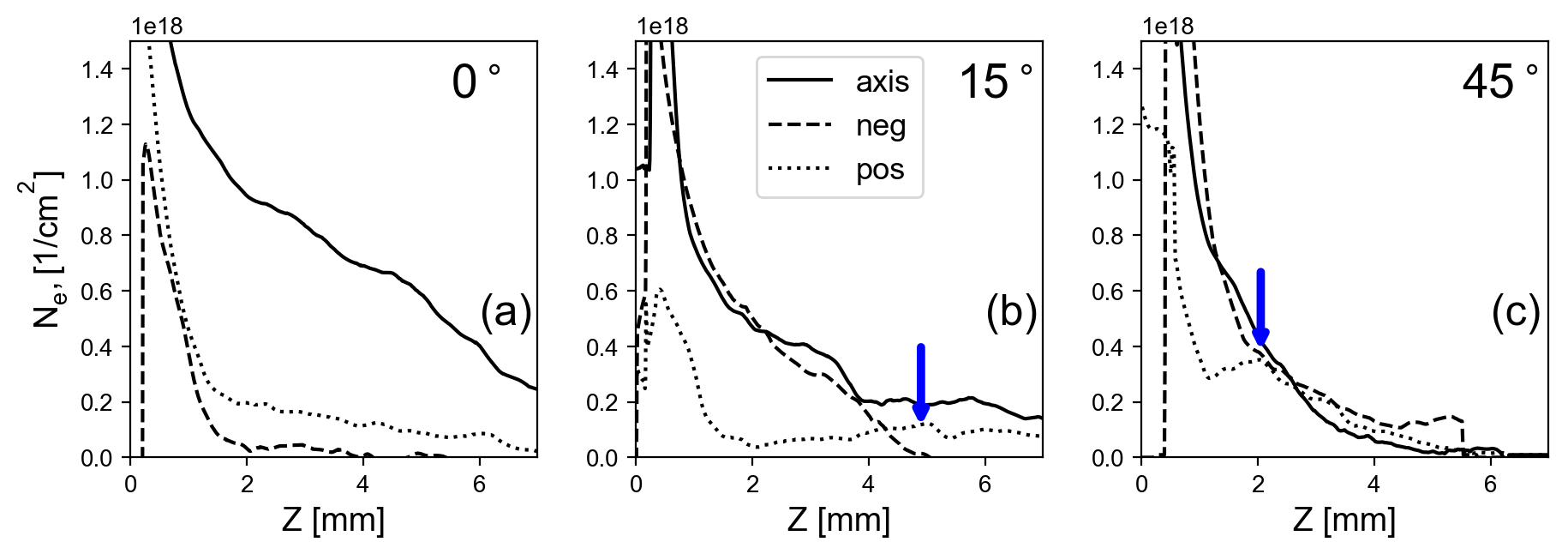}
    \caption{Profiles of the line-integrated electron plasma density at 96 ns along the Z-axis (solid line), along $r = -0.92$ mm  post-shock wing (dashed line) and along $r = 1.17$ mm post-shock wing (dotted line). The various lines are directly represented in the images shown in Figure \ref{fig:interferometry 30T} (f),(i),(l).}
    \label{fig:profiles}
\end{figure}

The lineouts of the two-dimensional electron plasma density along the axis normal to the surface of the obstacle target for 0$^{\circ}$, 15$^{\circ}$ and 45$^{\circ}$ stream impact at 96 ns are presented in Fig. \ref{fig:profiles}. We choose to make lineouts along the target normal to characterise how the rotation affects the development of the shoulders of the envelope close to the target, around the column. The positive lineout crosses the incoming stream, but it is not dense compared to the ejecta in this region. They demonstrate characteristic features of propagating plasma  (see dashed and dot-dashed lines in Fig. \ref{fig:interferometry 30T} (f),(i),(l)). The zero points of the axes ($z = 0$, $r = 0$)  corresponds to the stream tip arrival on the surface of the obstacle. The Z-axis corresponds to the axis of symmetry for normal stream incidence case (Fig. \ref{fig:interferometry 30T} (f)). Off-axis slices along Z correspond to the positions of the symmetrical edges of the ‘wings’ of the post-shock front for the normal impact (Fig. \ref{fig:interferometry 30T} (f)). For slanted impacts (15$^{\circ}$ and 45$^{\circ}$) such ‘wings’ evolve strongly asymmetrically and become less distinguishable with time.

For the normal impact case, the density distribution along the zero axis is much higher compared to the density profiles along the wings of the post-shocked plasma, which are monotonously decreasing and have both quite similar values. 

For the tilted cases, the positive off-axis density profile (dotted lines) exhibit a noticeable ‘hump’ in the profile (see the arrows pointing to it in Fig. \ref{fig:profiles}) which appears closer to the obstacle target with increasing tilt (about 5.0 mm from the target for 15$^{\circ}$ (Fig. \ref{fig:profiles} (b)) and 2.0 mm for 45$^{\circ}$ (Fig. \ref{fig:profiles} (c)). 

The negative off-axis density profiles for all cases decrease monotonously. For the tilted cases, the positive and the negative off-axis density-profiles notably differ from each other comparing to the symmetrical normal impact case. The on-axis profile in Fig. \ref{fig:profiles} decreases strongly for the slanted case due to the weaker shock structure and lack of plasma column confinement, while the negative-off axis lineout strongly increases, which directly evidences plasma leakage in the negative direction.

The zero axis density profiles for normal and tilted cases demonstrate also a ‘plateau and sharp decrease’ structure of the density front at a distance from the obstacle surface. Such structure becomes closer to the obstacle surface with increasing tilt angle (about 5.5 mm for 0$^{\circ}$ (Fig. \ref{fig:profiles} (a)), 4 mm for 15$^{\circ}$ (Fig. \ref{fig:profiles} (b)), 1.5 mm for 45$^{\circ}$ (Fig. \ref{fig:profiles} (c))). This quantitatively shows the increasing difficulty in refocusing matter on axis, following impact for increasingly slanted accretion columns. For the tilted cases the difference between the values of the on-axis density profile and that of the wings becomes clearly less pronounced and decreasing faster with increasing tilt angle. This is a direct representation of the much weaker shock and less confinement in the slanted case.

\begin{figure*}[htp]
    \centering
    \includegraphics[width=15 cm]{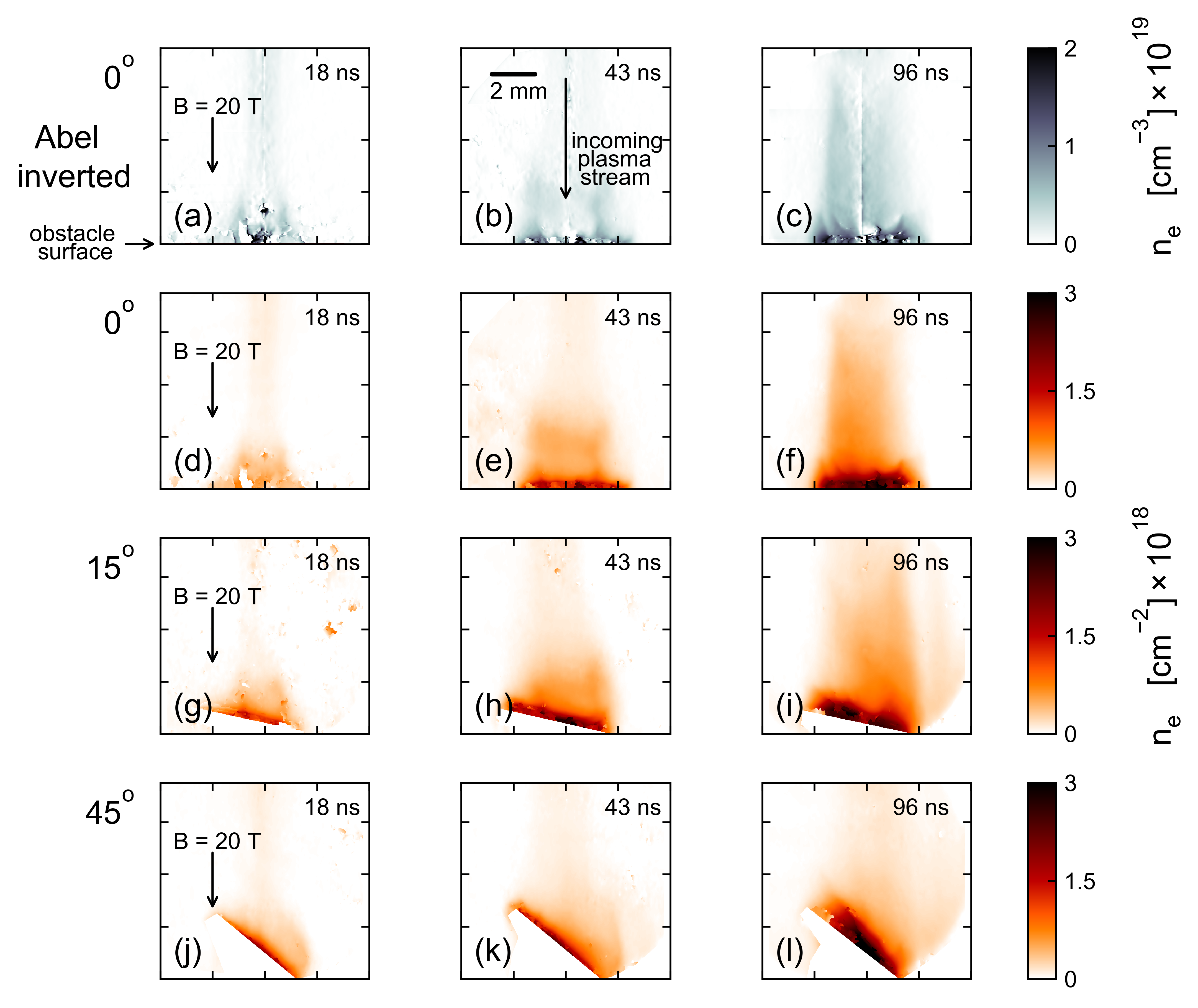}
    \caption{Evolution of plasma electron density with the same arrangement as for Fig. \ref{fig:interferometry 30T} but for lower $B = 20$ T.}
    \label{fig:interferometry 20T}
\end{figure*}

Reducing the magnetic field strength to 20 T, as shown in Fig. \ref{fig:interferometry 20T}, we observe similar effects, expect with wider overall structures.

\subsection{Visible self-emission of plasma}
In practice, the interferometry diagnostic is limited to probing density regions up to $2-5\times 10^{19}$ cm$^{-3}$. Due to refraction of the probe beam at distances closer than 1 mm from the target, surface interferometry cannot provide correct values of electron plasma density in this region. 
Self-emitted light is however able to escape from such dense plasma since the plasma is still optically thin. It is thus there that the SOP diagnostic is useful to give information about dense plasma regions close to the obstacle surface. The SOP diagnostic records, in the visible band (400-600 nm) and resolved in time, the emission of a thin ($2$ mm wide) slice of plasma along the Z-axis, centered at $r = 0$.  

\begin{figure*}[htp]
    \centering
    \includegraphics[width=18 cm]{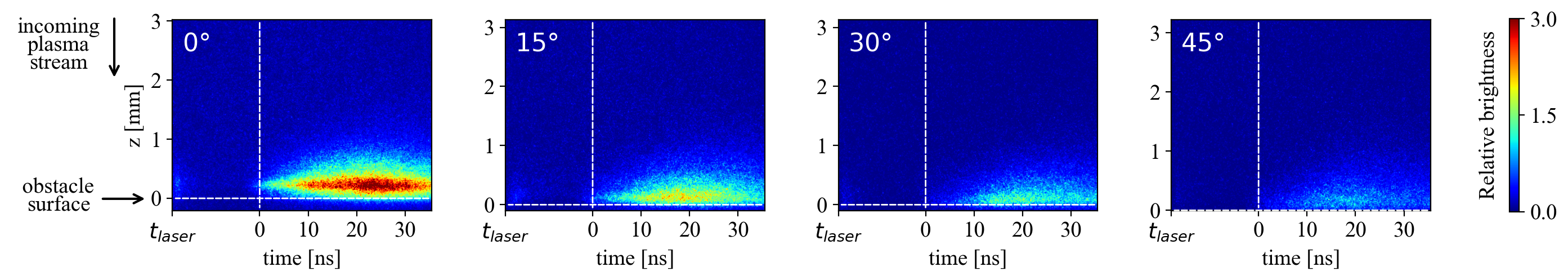}
    \caption{Visible self-emission of plasma measured by the SOP diagnostic for $B = 20$ T. The colorbar, which represents the emissivity in arbitrary units, is the same for all images.}
    \label{fig:SOP}
\end{figure*}

As shown in Fig. \ref{fig:SOP}, the SOP diagnostic can record emission of the plasma as close as 80 $\mu$m from the obstacle surface, i.e. in regions corresponding to densities going up to $10^{21}$ cm$^{-3}$, according to 3D simulations of the experiment (\citealt{2017SciA....3E0982R}). The SOP emission shown in Fig. \ref{fig:SOP} for various inclination of the obstacle target clearly shows that the density in the shocked region close to the obstacle surface is strongly reduced as the inclination of the obstacle is increased. Indeed, in this regime of optically thin plasma, we expect that the total emissivity of the plasma is proportional to the density of the emitting plasma. Here we present the SOP data for 20 T because of lack of 30 T SOP data, but since interferometry for both 20 T and 30 T cases represents almost the same features, comparative decreasing of plasma emission with increasing of the obstacle target tilt should be the same in the two cases.

Suchwise with the help of the interferometry and the SOP we have complementary diagnostics for low density and high density electron plasma, representing lateral ejection of material and subsequent less axial confinement of plasma under oblique impact.

\subsection{X-ray self-emission of plasma}
\subsubsection{Experimental results}
Similarly to the SOP, the VSG diagnostic records the plasma emissivity, but this time in a broadband X-ray range (0.4-2 keV). It also resolves the plasma emission along the Z axis, but the signal is integrated along the radial (r) axis, as well as integrated in time. However, the emission is spectrally resolved in the range of sensitivity. We used the same setup as described in \citealt{Kita:83} with the concave grating with an incidence angle of 87$^{\circ}$ and nominal groove number of 1200 grooves/mm. The detector used was a FuijiFilm image plate type TR. In front of the image plate was a sheet of aluminized plastic to serve as a light-tight filter and also created absorption edges that were used to calibrate the energy dispersion. For the deconvolution of the VSG data, the dependencies of the image plate sensitivity (\citealt{10.1117/12.2024889}) and of the reflectivity of the grating (\citealt{doi:10.1063/1.3495790}) on the energy of photons were both taken into an account.

\begin{figure*}[htp]
    \centering
    \includegraphics[width=16cm]{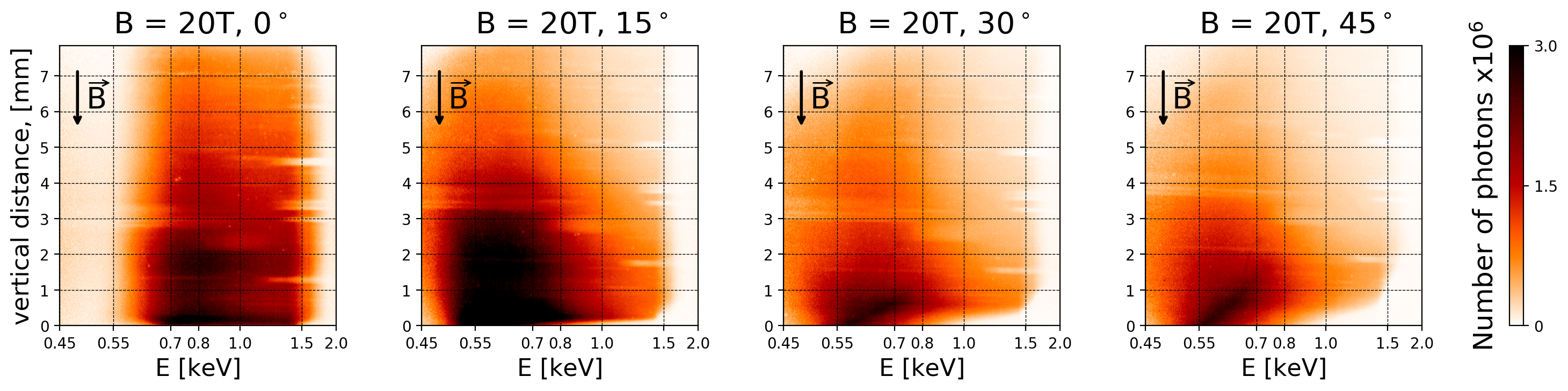}
    \includegraphics[width=16cm]{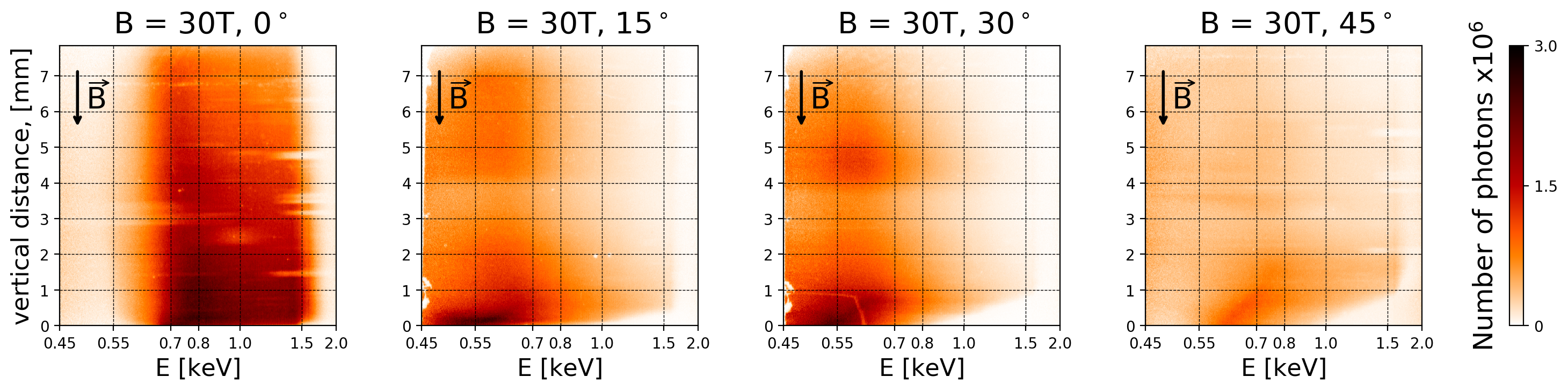}
    \caption{Energy spectra of the plasma self-emission recorded in the soft X-ray, as measured by the VSG diagnostic for $B = 20$ T (top row) and $B = 30$ T (bottom row; in the 30T case for 15$^{\circ}$ and 30$^{\circ}$ the lighter band around $z = 3-4$ mm is an artefact due to a local loss of sensitivity of the detector which was used for these shots). These are raw data corrected for the detector (image plate) response to the X-rays and for the grating reflectivity, which is dependant on the X-ray energy (see text).}
    \label{fig:VSG}
\end{figure*}

The VSG emission shown in Fig. \ref{fig:VSG} for various inclination of the obstacle target clearly shows, complementary to the SOP diagnostics, that: (1) the confinement of the plasma along the Z axis strongly decreases as the incoming stream is increasingly slanted. This is evidenced by the strong decrease of the amplitude of the emission along Z as the obstacle inclination is increased. Also, (2) the plasma near the obstacle is colder as the incoming stream is increasingly slanted, as evidenced by the progressive drift of the signal toward lower photon energies. This is consistent with a lesser capability of the plasma to be confined due to the increased angle between the magnetic field and the incoming stream axis.

Note that the VSG is equipped with a vertical slit (perpendicular to the incoming stream axis, Z, which, in the laboratory, is horizontal) in order to allow spatial resolution along the axis Z. The grating has its lines in the horizontal direction. Hence we have horizontal spatial resolution and vertical spectral resolution on the VSG detector. But since we used a slit instead of a pinhole at the VSG entrance, the spectrum in the vertical axis is not purely spectral and is mixed with space (various emitting points in the plasma at different height (along the axis r) will appear at various heights on the IP). Another effect of using a slit is that the tilted targets appear also tilted in the IP detector, as is apparent in Fig. \ref{fig:VSG}. The mixing of space and spectrum induces some smearing on the recorded spectra, such that we do not have enough spectral resolution to resolve exact transition lines. However, the observed spectral changes, when tilting the obstacle, have to be compared to the size of the smearing induced by the source height. In real space, the spatial width of the plasma source in the vertical direction is 1-2 mm large, which transposes in the same distance on the detector as there is a magnification of one through the VSG. This has to be compared to the typical extend of the spectral features shown in Fig. \ref{fig:VSG}, which are of the order of 6-8 mm on the detector, i.e. quite large in comparison to the smearing induced by the source size.

\begin{figure}[htp]
    \centering
    \includegraphics[width=8cm]{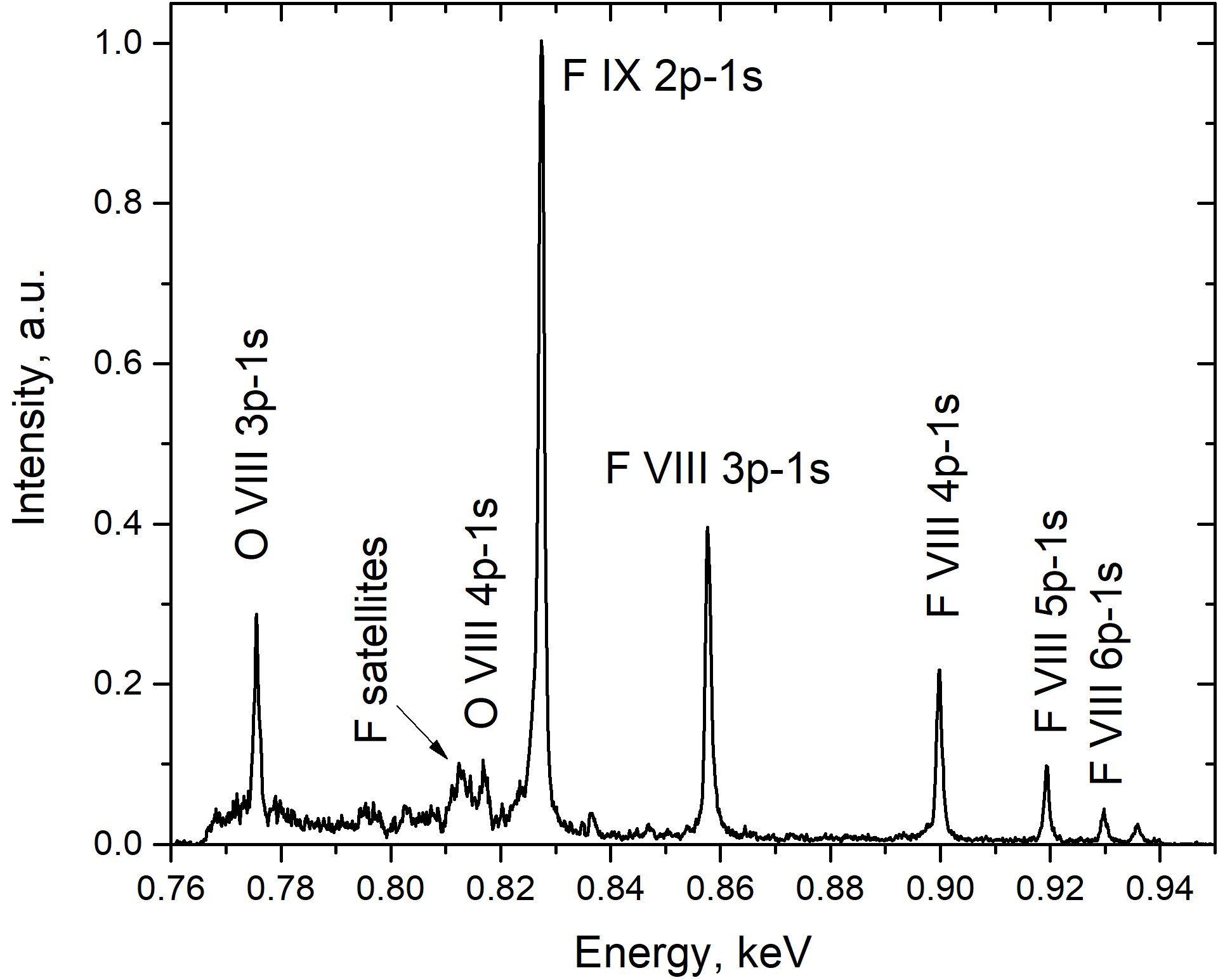}
    \caption{Typical spectrum measured by the FSSR spectrometer for the case of normal incidence. The FSSR is placed at 90$^{\circ}$ to the target normal axis with a spatial resolution along the plasma expansion axis. The observation of the Oxygen H-like transition confirms that the electron temperature close to the target is high.}
    \label{fig:FSSR}
\end{figure}

The FSSR, which has a much better spectral resolution, but over a smaller bandwidth, allows us to resolve individual lines. A spectrum recorded by the FSSR is shown in Fig. \ref{fig:FSSR} for the emission close to the obstacle target surface (around 0.25 mm) for the 0 degrees (normal incidence) case. We can use such spectrum to measure the electron temperature and density in a similar way as described in \citealt{2017SciA....3E0982R, Filippov19}. The plasma parameters of the hot cocoon surrounding the colder core of the impact of the incoming stream, as well as that of the core are respectively deduced to be 350 eV/$3\times 10^{18}$ cm$^{-3}$ and 50 eV/$3\times 10^{19}$ cm$^{-3}$. Close to the obstacle we also note the observation of oxygen lines in the spectrum, having a high quantum number n, thus confirming that the plasma temperature is high. However, the relative abundance of O in the plasma, due to the mixing of the incoming stream plasma and of that of the obstacle, is unknown. 
To further the analysis of the plasma parameters for different obstacle orientations, we turn to using the VSG spectrometer, despite its low spectral sensitivity, taking advantage of its capability to record emissivity over a larger spectral range. This is what is discussed in the next Section.

\subsubsection{Numerical modeling of the laboratory plasma X-ray emissivity}
We have modeled the plasma emissivity over a broad spectral range in order to be able to infer the plasma temperature from the VSG spectra. Since the recorded VSG spectra are integrated along the radial axis as well as in time, many different plasma conditions will have therefore contributed to the emissivity. However, even with insufficient spectral resolution of the VSG spectrometer, the comparison with numerical synthetic spectra, as discussed below, may provide information about the maximum temperatures reached in the plasma plume. With that purpose, we carried out numerical simulations of synthetic spectra of CF$_2$ plasmas using the MIXKIP/RAPCAL codes (\cite{Rodriguez08,Espinosa17}). For the calculations, an electron density of $10^{19}$ cm$^{-3}$ and a plasma length of $1.5$ mm were used. According to Fig.\ref{fig:interferometry 30T} both average values are reasonable to model the plasma close to the target surface ($z=1.5$ mm). Moreover, since there is no time resolution, the synthetic spectrum for a given electron temperature was obtained by adding those calculated for temperatures between 5 eV and the given temperature. To illustrate the results, the situation in which the strength of the magnetic field is 30 T and $z=1.5$ mm was selected.

\begin{figure}[htp]    
\centering
\includegraphics[width=8.5 cm]{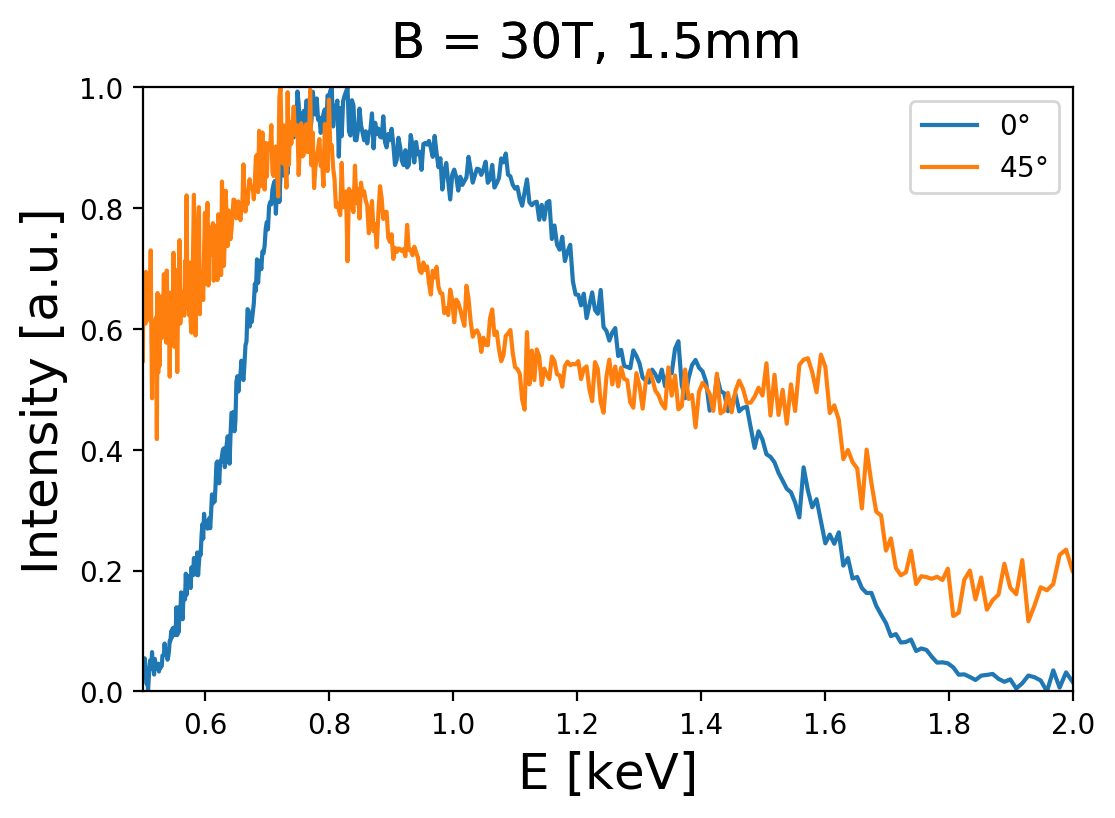}
\caption{Spectral lineouts of the experimental X-ray emissivity spectra shown in Fig. \ref{fig:VSG}. The lineouts correspond to the images shown in the bottom row of Fig. \ref{fig:VSG}, i.e. to the case of $B = 30$ T, for 0$^{\circ}$ and 45$^{\circ}$ angles. The lineouts are taken at the position $z = 1.5$ mm. }
\label{lineout30T}
\end{figure}

Furthermore, for the comparison with the experiment, we focused on the situation of the normal impact case and of the 45$^{\circ}$ oblique stream incidence. In Fig. \ref{lineout30T} are presented the experimental lineouts corresponding to the VSG spectra recorded in these two cases, i.e. these are the lineouts at the position $z = 1.5$ mm of the raw data shown in Fig. \ref{fig:VSG}, in the bottom row: the left most (normal incidence) and right most (45$^{\circ}$ oblique stream incidence) images. Note that here we will discuss, in the light of the numerical modeling detailed below, only the features of these experimental spectra below 1 keV photon energy since it is likely that for higher photon energies, the emission could be originating from recombination continuum.

\begin{figure}[htp]    
\centering
\includegraphics[width=6.7 cm]{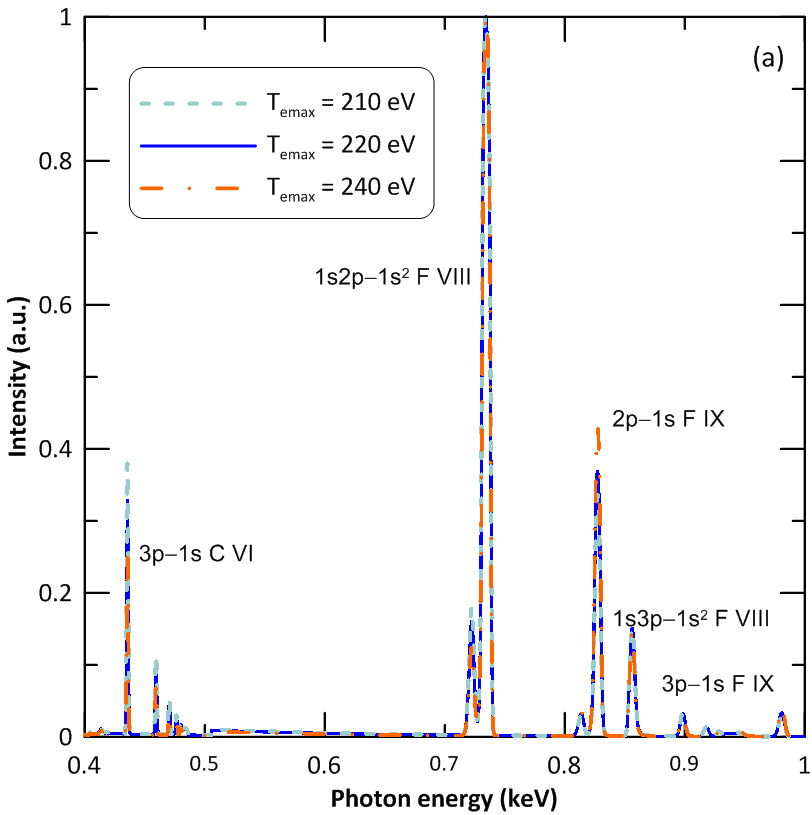}
\includegraphics[width=6.7 cm]{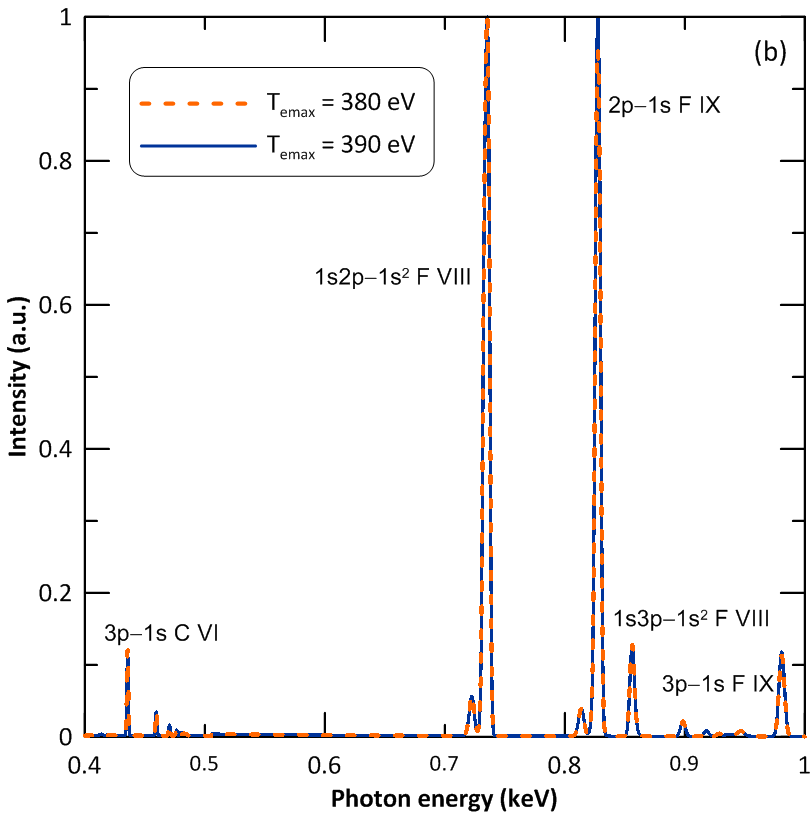}
\includegraphics[width=6.7 cm]{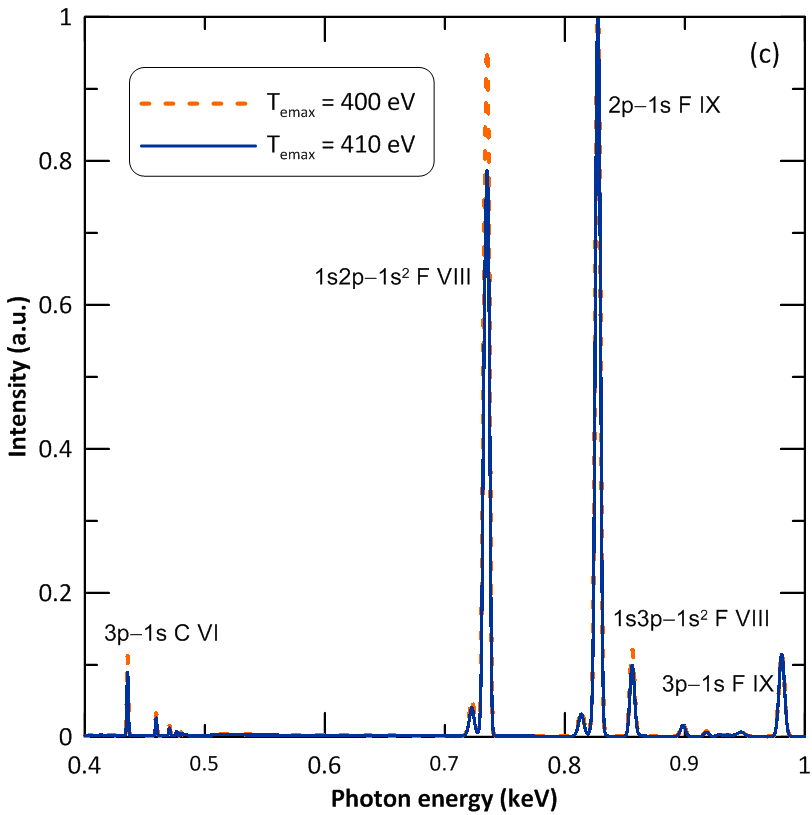}
\caption{Synthetic spectra of CF${_2}$ plasma obtained for different electron temperatures and for an electron density of $10^{19}$ cm$^{-3}$ and a plasma length of $1.5$ mm. The photon energy range represented is that in which the transition lines used in the analysis contribute.}
\label{synthetic}
\end{figure}

The emission lines that can be modelled for various plasma temperatures are presented in Fig.\ref{synthetic}. Three ranges of temperatures are modelled: 210-240 eV (Fig.\ref{synthetic} (a)), 380-390 eV (Fig.\ref{synthetic} (b)), and 400-410 eV (Fig.\ref{synthetic} (c)). One can observe that between these three cases, the ratio of the various groups of lines varies. 

%\textcolor{blue}{The spectral location of the transition lines of F and C ions used to estimate the plasma temperatures has also been indicated in Figure \ref{lineout30T}.}

 For the lowest considered temperatures, 210-240 eV (Fig.\ref{synthetic} (a)), one can see (1) that the relative weight of the spectral intensities corresponding to the $2p-1s$ transitions of F IX ion is quite lower than that of the $1s2p-1s^{2}$ transitions of F VIII ion. This is contrary to what happens at higher temperatures. Still for the lowest considered temperatures, one observes (2) that the contributions provided by the transitions of the C VI ion below 0.5 keV photon energy range are of similar intensity as that of the $2p-1s$ transitions of F IX ion (around 0.83 keV photon energy). These two observations are best matching the variation of the spectral intensity observed for the 45$^{\circ}$ case. This suggests that the plasma temperature is low in this case. In the normal incidence case, the situation seems however different and would suggest a higher temperature. Indeed, in that case, Fig.\ref{lineout30T} shows (1) that the spectral intensity around 0.5 keV is much smaller in comparison to that around 0.75 keV. Moreover, contrary to the 45$^{\circ}$ case, (2) the spectral intensities from 0.75 to 0.85 keV are quite similar. All this is best matched by the case of the hottest considered plasma in the simulations, i.e. that of Fig.\ref{synthetic} (c). Indeed, in that case, the intensities around 0.8 keV, corresponding to the $2p-1s$ transitions of F IX ion and the $1s2p-1s^{2}$ transitions of F VIII ion, have similar amplitudes, and the contribution of the transitions of the C VI ion is low. Therefore, this suggests that electron temperatures around 390-410 eV are reached for the overall plasma plume in the normal incidence case.

%for the normal impact case.which indicates that the maximum temperatures reached in the latter case are lower.

   %Furthermore, in this range of temperatures the contributions provided by the numerical simulations of the transitions of the C VI ion in the 430-480 eV photon energy range are small, which agrees with the experimental results.
%Fig.\ref{lineout30T} shows that the spectral intensities corresponding to the $3p-1s$ transitions of the C VI are slightly weaker than $2p-1s$ transitions of the F IX ion. The numerical simulations indicate that this behavior is not obtained for temperatures lower than 220 eV.
%Fig.\ref{synthetic}(a) also shows that for temperatures of 240 eV and higher, the relative differences of the strengths of these transitions begin to be more noticeable. Hence, we may conclude that the range of maximum temperatures reached in this case would be 220-240 eV.

%\textcolor{blue}{For the tilt angle of 45$^{\circ}$, Fig.\ref{lineout30T} shows that the contribution due to the $np-1s$ transitions of the C VI ion significantly increases with respect to the normal impact case,}

%\textcolor{blue}{Fig.\ref{lineout30T} shows that the $3p-1s$ transitions of the C VI are slightly weaker than $2p-1s$ transitions of the F IX ion. The numerical simulations indicate that this behavior is not obtained for temperatures lower than 220 eV (Fig.\ref{synthetic}(a)). On the other hand, Fig.\ref{synthetic}(a) also shows that for temperatures of 240 eV and higher, the relative differences of the strengths of these transitions begin to be more noticeable. Hence, we may conclude that the range of maximum temperatures reached in this case would be 220-240 eV.}

%Furthermore, a great feature in the intensity is detected in the 650-700 eV photon energy range which do not correspond to transitions of C or F.
%...and the $1snp-1s^{2}$ (with $n=3-5$) of the O VII ion and the $2p-1s$ transitions of the O VIII ion have transition energies that lie in that range of photon energies and, therefore, these transitions could explain that feature.
%Furthermore, the numerical simulations also provide that in this range of electron temperatures, the most abundant ions of O are O VII and O VIII, which also agrees with the fact that their contributions are recorded in the spectrum.

\section{Modelling accretion impacts in young stars}
\label{sec:mod.result}
\subsection{The MHD model}
The results of the experiment were compared with MHD simulations describing  the final propagation of an accretion stream through the atmosphere of a CTTS and its impact onto the chromosphere of the young stellar object. To this end, we adopted the model of \cite{2010A&A...510A..71O} which assumes a stream of plasma downflowing along magnetic field lines that link the circumstellar disk to the stellar surface. The magnetic field is assumed to be uniform but, at variance with the model of \cite{2010A&A...510A..71O}, the field can have an angle smaller than 90$^{\circ}$ with respect to the stellar surface. The computational domain includes only the portion
of the stellar atmosphere where the impact occurs. The model assumes
a fully ionized plasma with a ratio of specific heats $\gamma =
5/3$, and accounts for the effects of gravity, radiative cooling,
and thermal conduction (including the effects of heat flux saturation).
The stream impact is modeled by numerically solving the time-dependent
MHD equations of mass, momentum, and energy conservation in
non-dimensional conservative form:

\begin{equation}
\frac{\partial \rho}{\partial t} + \nabla \cdot (\rho \vec{u}) = 0~,
\end{equation}

\begin{equation}
\frac{\partial \rho \vec{u}}{\partial t} + \nabla \cdot (\rho
\vec{u}\vec{u}-\vec{B}\vec{B}) + \nabla P_* = \rho \vec{g}~,
\end{equation}

\begin{equation}
\frac{\partial \rho E}{\partial t} +\nabla\cdot [\vec{u}(\rho
E+P_*) -\vec{B}(\vec{u}\cdot \vec{B})] = \rho \vec{u}\cdot
\vec{g} -\nabla\cdot \vec{F}_{\rm c} -n_{\rm e} n_{\rm H}
\Lambda(T)~,
\end{equation}

\begin{equation}
\frac{\partial \vec{B}}{\partial t} +\nabla
\cdot(\vec{u}\vec{B}-\vec{B}\vec{u}) = 0~,
\end{equation}

\noindent
where

\[
P_* = P + \frac{B^2}{2}~,~~~~~~~~~~~~~
E = \epsilon +\frac{1}{2} u^2+\frac{1}{2}\frac{B^2}{\rho}~,
\]

\noindent
are the total pressure, and the total gas energy (internal energy,
$\epsilon$, kinetic energy, and magnetic energy) respectively, $t$ is the
time, $\rho = \mu m_H n_{\rm H}$ is the mass density, $\mu = 1.28$ is the
mean atomic mass (assuming metal abundances of 0.5 of the solar values;
\citealt{Anders1989GeCoA}), $m_H$ is the mass of the hydrogen atom, $n_{\rm
H}$ is the hydrogen number density, $\vec{u}$ is the gas velocity, $g$
is the gravity, $T$ is the temperature, $\vec{B}$ is the magnetic field,
$\vec{F}_{\rm c}$ is the conductive flux, and $\Lambda(T)$ represents
the optically thin radiative losses per unit emission measure derived
with the PINTofALE spectral code (\citealt{Kashyap2000BASI}) with the APED
atomic line database (\citealt{Smith2001ApJ}), assuming the same metal
abundances as in \cite{2010A&A...510A..71O} (as deduced from X-ray observations of CTTSs;
\citealt{Telleschi2007A&Ab}). We use the ideal gas law, $P=(\gamma-1)
\rho \epsilon$.

The thermal conductivity is highly anisotropic due to the presence
of the stellar magnetic field: it is strongly reduced in the direction
transverse to the field (e.g. \citealt{spi62}). Furthermore, the
model includes the classical and the saturated conduction regime.
Thus, we treat the thermal flux as in \cite{obr08}, by splitting
the thermal flux into two components, along and across the magnetic
field lines, ${\bf F}\rs{c} = F_{\parallel}~{\bf i}+F_{\perp}~{\bf
j}$, where

\begin{equation}
\begin{array}{l}\displaystyle
F_{\parallel} = \left(\frac{1}{[q_{\rm spi}]_{\parallel}}+
                \frac{1}{[q_{\rm sat}]_{\parallel}}\right)^{-1}~,
\\ \\ \displaystyle
F_{\perp} = \left(\frac{1}{[q_{\rm spi}]_{\perp}}+
               \frac{1}{[q_{\rm sat}]_{\perp}}\right)^{-1}~.
\end{array}
\label{cond}
\end{equation}

\noindent
$[q_{\rm spi}]_{\parallel}$ and $[q_{\rm spi}]_{\perp}$ are the
classical conductive flux along and across the magnetic field lines following (\citealt{spi62}), and $[q_{\rm sat}]_{\parallel}$ and $[q_{\rm sat}]_{\perp}$ are the saturated flux along and across the magnetic field lines following (\citealt{cm77}). Eqs.~\ref{cond} allow for a smooth transition between the classical and saturated conduction regime
(see \citealt{1993ApJ...404..625D, obr08}).

The gravity is calculated assuming the star mass $M=1.2 M_{\sun}$
and the star radius $R=1.3 R_{\sun}$ which are appropriate for the
CTTS MP~Mus (see \citealt{amp07}). In the present calculations,
however, the effects of gravity are considered only in the chromosphere,
and are not included in the stellar corona and the stream, at
variance with the model of \cite{2010A&A...510A..71O}. This was
done to allow for a more direct comparison of the model results
with those of our laboratory experiment where the gravity effects
can be considered to be negligible. The effects of gravity are
considered in the stellar chromosphere to have a realistic density
stratification there.

Initially the stellar atmosphere is unperturbed and in magneto-static
equilibrium. The vertical profiles of mass density and temperature
from the base of the transition region ($T=10^4$~K) to the corona
are calculated by using the wind model of \cite{1996JGR...10124443O}
adapted to the conditions of a CTTS. Thus the stellar atmosphere
consists of a hot (maximum temperature $\approx 10^6$ K) and tenuous
($n_{\rm H} \approx 2\times 10^8$ cm$^{-3}$) corona linked through
a steep transition region to an isothermal chromosphere that we
model with temperature $10^4$~K and thickness $8.5\times 10^8$~cm.

At the beginning of the simulation, the accretion stream enters
into the spatial domain from the upper boundary. As reference case,
we considered a stream with density $n_{\rm str0} = 10^{11}$~cm$^{-3}$
and velocity $u_{\rm str0} = -500$~km~s$^{-1}$ compatible with those
derived from the analysis of X-ray spectra of MP Mus (\citealt{amp07}).
The stream pressure is determined by the pressure balance across
the stream lateral boundary. The unperturbed ambient magnetic field
is assumed to be uniform with strength $|\vec{B}\rs{0}| = 10$~G and
inclined with respect to the stellar surface with an angle $\theta
= 80^{\rm o}$.  Then we considered additional simulations in which
we alternatively vary the stream density, and the strength and
inclination of the stellar magnetic field, around the reference
values. Since the stream reproduced in the laboratory
experiment has characteristics similar to those of the stream
analyzed by \cite{2017SciA....3E0982R}, we considered magnetic field
strengths that lead to a plasma $\beta$ in the post-shock region
similar to that found in \cite{2017SciA....3E0982R}, ranging between
1 and 100. Table~\ref{tab1} reports a summary of the simulations
performed, where $n_{\rm str0}$ and $u_{\rm str0}$ are the initial
density and velocity of the stream, respectively, $|\vec{B}\rs{0}|$
is the initial magnetic field strength, and $\theta$ is the inclination
angle of the stream with respect to the stellar surface.

We performed the calculations with PLUTO, a modular Godunov-type
code for astrophysical plasmas, intended mainly for astrophysical
applications and high Mach number flows in multiple spatial dimensions
(\citealt{mbm07}). The MHD equations are solved using the MHD module
available in PLUTO, configured to compute intercell fluxes with the
Harten-Lax-van Leer Discontinuities (HLLD) approximate Riemann
solver, while second order in time is achieved using a Runge-Kutta
scheme. A monotonized central difference limiter for the primitive
variables is used. The evolution of the magnetic field is carried
out adopting the constrained transport approach
(\citealt{1999JCoPh.149..270B}) that maintains the solenoidal
condition ($\nabla\cdot\vec{B}=0$) at machine accuracy. The optically
thin radiative losses are included in a fractional step formalism
(\citealt{mbm07}), which preserves the $2^{nd}$ time accuracy, as
the advection and source steps are at least of the $2^{nd}$ order
accurate. The radiative losses $\Lambda$ values are computed at the
temperature of interest using a table lookup/interpolation method.
The thermal conduction is treated by adopting the super-time-stepping
technique (\citealt{aag96}) which is very effective to speed up
explicit time-stepping schemes for parabolic problems.

We solved the MHD equations using cartesian coordinates in the plane
$(x,z)$. The coordinate system is oriented in such a way that the
stellar surface lies on the $x$-axis and the normal to the stellar
surface is along the $z$-axis. The stream axis is inclined with
respect to the stellar surface and the $x$-axis by the angle $\theta$.
The mesh extends between $-2.2\times 10^{10}$~cm and $1.1\times
10^{10}$~cm in the $x$-direction and between $-1.0\times 10^9$~cm
and $1.4 \times 10^{10}$~cm in the $z$-direction. The transition
region between the chromosphere and the corona is located at $z=0$~cm.
The grid along the $x$-axis is uniform and made of $N\rs{r} = 256$
points with a resolution of $\Delta x \approx 1.3\times 10^8$~cm;
the grid along the $z$-axis is nonuniform with mesh size increasing
with $z$. In such a way, the $z$-grid has the highest spatial
resolution closer to the stellar chromosphere, thus allowing an
accurate description of the steep temperature gradient of the
transition region and the evolution of post-shock material, resulting
from the impact of the accretion stream with the stellar chromosphere.
More specifically, the $z$-grid is made of $N\rs{z} = 512$ points
and consists of a uniform grid patch with 256 points and a maximum
resolution of $\Delta z \approx 10^7$~cm that covers the chromosphere
and the upper stellar atmosphere up to the height of $\approx
1.8\times 10^9$~cm and a stretched grid patch for $z > 1.8\times
10^9$~cm with 256 points and a mesh size increasing with $z$, which
leads to a minimum resolution at the upper boundary of $\Delta z
\approx 2.8\times 10^8$~cm.

The boundary conditions are free outflow\footnote{Set zero gradients
across the boundary.} at $x = -2.2 \times 10^{10}$~cm and $x = 1.1
\times 10^{10}$~cm, fixed boundary conditions at $z = -1.0\times
10^9$~cm (imposing zero material and heat flux across the boundary),
and a constant inflow in the upper boundary at $z = 1.4 \times
10^{10}$ cm.

\begin{table}
\centering
\caption{Parameters for the MHD models of accretion impacts}
\label{tab1}
\begin{tabular}{lcccc}
\hline
\hline
Model    &  $|\vec{B}\rs{0}|$ & $n\rs{str0}$ & $u\rs{str0}$ & $\theta$  \\
abbreviation & [G] & [$10^{11}$ cm$^{-3}$] & [km s$^{-1}$]  &  \\
\hline
D1e11-B10-A10  & 10  & 1    & 500 & 10 \\
D1e11-B10-A20  & 10  & 1    & 500 & 20 \\
D1e11-B50-A20  & 50  & 1    & 500 & 20 \\
D5e10-B10-A20  & 10  & 0.5  & 500 & 20 \\
D5e10-B50-A20  & 50  & 0.5  & 500 & 20 \\
D5e10-B30-A15  & 30  & 0.5  & 500 & 15 \\
D5e10-B30-A45  & 30  & 0.5  & 500 & 45 \\
\hline
\hline
\end{tabular}
\end{table}

\begin{figure*}[htp]
    \centering
    \includegraphics[width=15 cm]{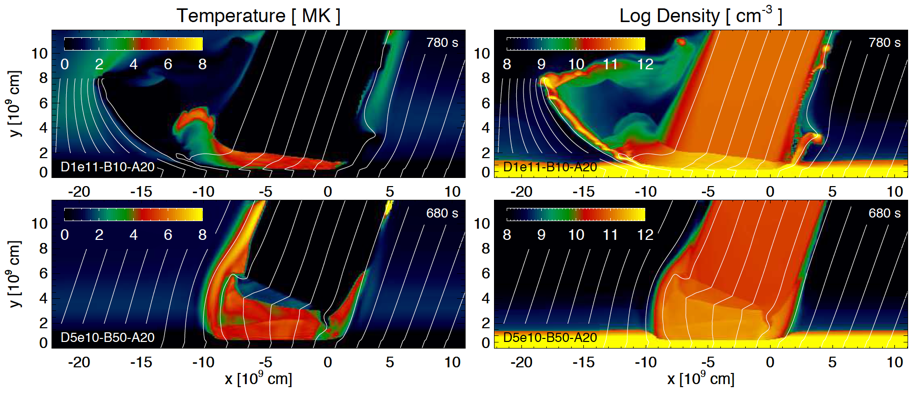}
    \caption{Spatial distributions of temperature (left panels) and density (right panels in log scale) at the labeled times for a model with high plasma $\beta$ (run D1e11-B10-A20; upper panels) and a model with low plasma $\beta$ (run D5e10-B50-A20; lower panels). The white lines mark magnetic field lines.}
    \label{fig:temp & dens distribution}
\end{figure*}

\subsection{Dynamics of the stream impact}
We found that the evolution of the stream impact is similar to that described in \cite{2010A&A...510A..71O} when the plasma $\beta$ is high (and the shock-heated plasma is poorly confined by the magnetic field) and to that described in \cite{refId0orl} when $\beta$ is low (and the post-shock plasma is well confined by the ambient field). The main difference with previous models is that now the stream hits the stellar surface with an incidence angle larger than 0$^{\circ}$ and this introduces a marked asymmetry of the density and temperature distributions in the post-shock plasma. Fig. \ref{fig:temp & dens distribution} shows 2D maps of temperature and density spatial distribution for two models characterized by high or low plasma $\beta$.  Movies showing the complete evolution of temperature and density in log scale for all the models are provided as online material. As in previous analogous simulations, the accreting material downflows along the magnetic field lines and hits the stellar chromosphere at $t \sim 250 s$. Then the stream gradually sinks into the chromosphere and it stops sinking when, locally, the thermal pressure of the chromosphere equals the ram pressure of the stream. At this time, a shock develops at the base of the accretion column and propagates upward through the stream, heating the accreting material up to temperatures of few millions degrees (thus contributing to emission mainly in the X-ray band). This newly formed hot slab of plasma is partially rooted in the chromosphere, so that the X-ray emitting plasma is buried under a column of optically thick material and is partially absorbed (e.g. \citealt{2013Sci...341..251R,Bonito_2014, 2017SciA....3E0982R, refId0cos, refId0col}).

Due to the high plasma $\beta$ value in our reference case (upper panels of Fig. \ref{fig:temp & dens distribution}), the dense hot plasma in the slab causes a pressure-driven flow which ejects part of the accreted material sideways. Due to the stream inclination with respect to the stellar surface, the flow velocity has a component, $v\rs{perp}$, perpendicular to the stellar surface. As a result, the effect of impact at the border of the stream is different on the two sides: the outflow is not symmetric around the stream and is more fast and pronounced in the direction of $v\rs{perp}$. This is particularly true in the case of models with high plasma $\beta$ (D1e11-B10-A20, D1e11-B10-A10, D5e10-B10-A20). In these cases, the outflow strongly perturbs the surrounding stellar atmosphere and drags the magnetic field trapped at the head of the escaped material, leading to a continuous increase of the magnetic pressure and field tension there (see upper panels of Fig. \ref{fig:temp & dens distribution}).  In general, due to the presence of the magnetic field, the escaped material does not flow freely but  is redirected upwards and, possibly, is pushed back on the stream by the magnetic field tension. In run D5e10-B50-A20, the magnetic field is strong enough to force the escaped material to plunge into the stream and to significantly perturb it (see lower panels of Fig. \ref{fig:temp & dens distribution}). On the side of the stream opposite to the direction of $v\rs{perp}$, the field  is strong enough to keep the outflow close to the stream in all the simulations explored. The outflowing plasma accumulates around the accetion column and gradually forms a sheath of turbulent material which envelops the stream.

In all our simulations, the hot slab is thermally unstable due to the radiative cooling. In fact, the continuous accumulation of accreting material at the base of the stream leads to an increase of density of the post-shock plasma and, as a consequence, to an increase of radiative losses (which depends on the square of density for optically thin plasmas). As a consequence, oscillations of the hot slab (and variability of the outflowing plasma) are induced by radiative cooling. In models with high plasma $\beta$, the variations appear chaotic without an evident periodicity, at least in the time lapse explored here. In models with low plasma $\beta$, quasi-periodic oscillations are observed although they are strongly perturbed by the sheath of turbulent material around the stream.

\begin{figure*}[htp]
    \centering
    \includegraphics[width=15 cm]{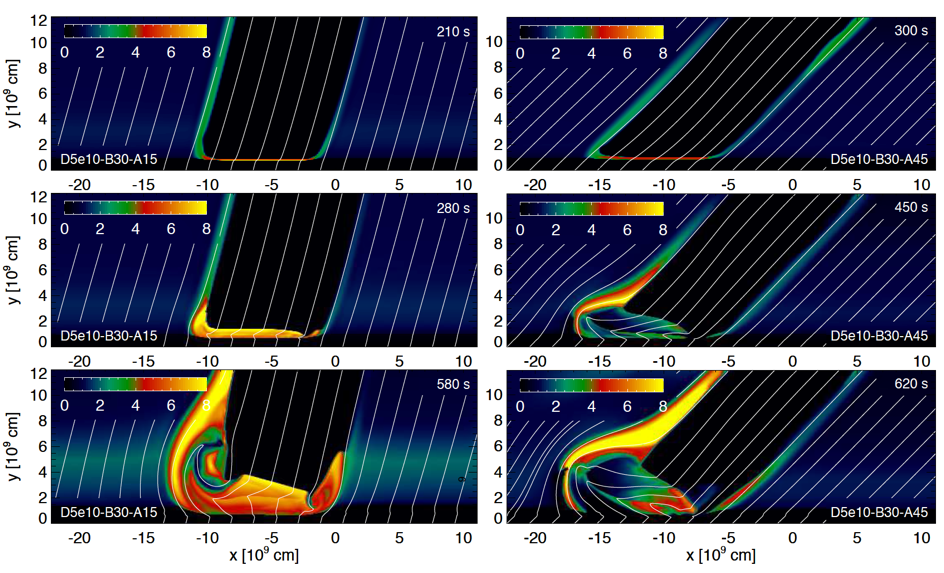}
    \caption{Evolution of temperature distribution at the labeled times for 15$^{\circ}$ (run D5e10-B30-A15; on the left) and 45$^{\circ}$ (run D5e10-B30-A45; on the right) oblique accretion stream incidence. The white lines mark magnetic field lines.}
    \label{fig: evolution}
\end{figure*}

\begin{figure*}[htp]
    \centering
    \includegraphics[width=15 cm]{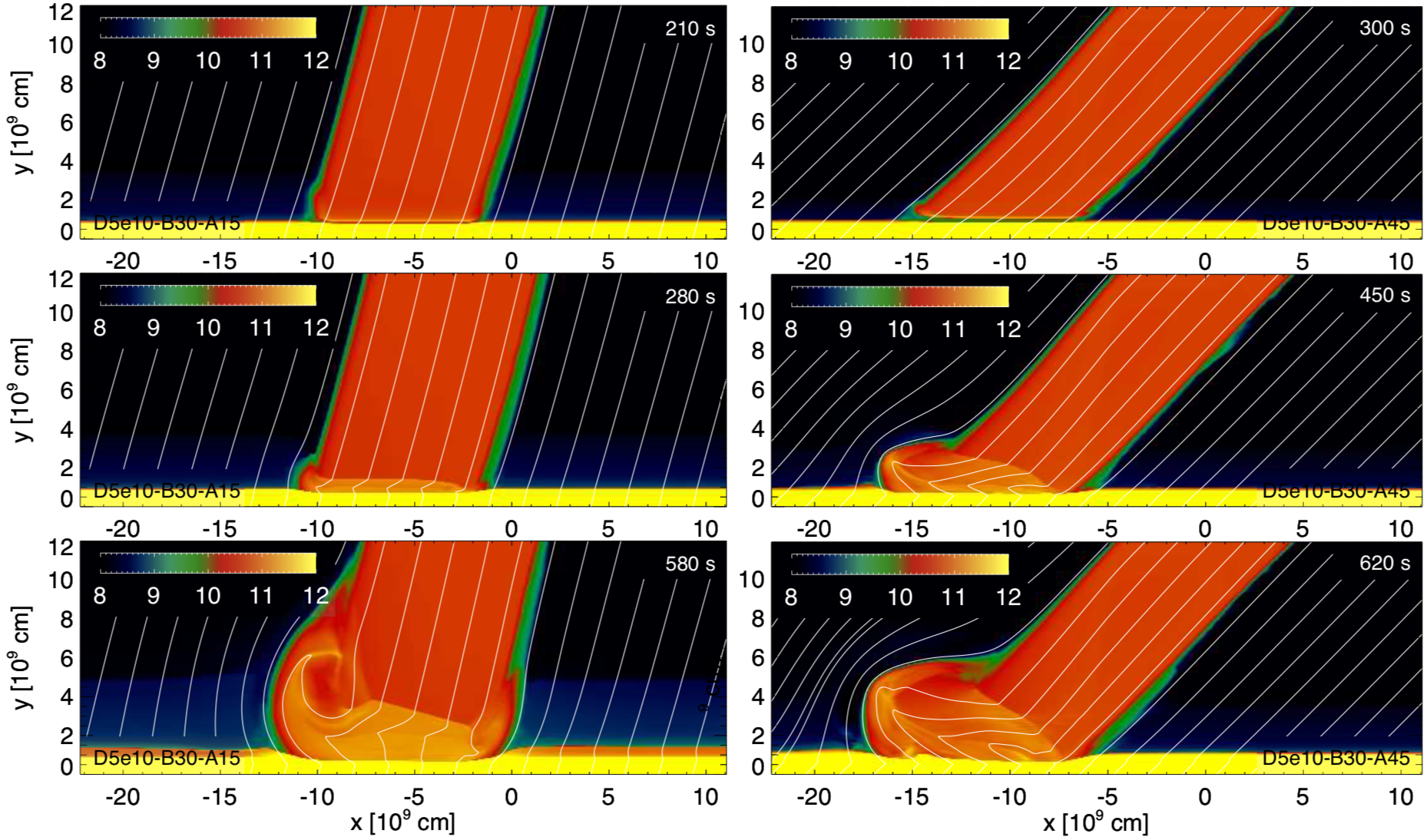}
    \caption{Spatial distributions of density in log scale at the labeled times for 15$^{\circ}$ (run D5e10-B30-A15; on the left) and 45$^{\circ}$ (run D5e10-B30-A45; on the right) oblique accretion stream incidence. The white lines mark magnetic field lines.}
    \label{fig:model_dens}
\end{figure*}

Among the models analyzed, those which roughly reproduce the morphology observed in the experiments assume a stream with a density $n_{\rm str0} = 5\times 10^{10}$~cm$^{-3}$ which propagates through a magnetized stellar atmosphere with magnetic field strength $|\vec{B}\rs{0}| = 30$~G. Fig. \ref{fig: evolution} and Fig. \ref{fig:model_dens} shows the distribution of temperature and density for two incidence angles, namely 15$^{\circ}$ and 45$^{\circ}$, the same of our experiments in Fig. \ref{fig:interferometry 30T}. In the first case the evolution is similar to that described above. In the extreme case of 45$^{\circ}$ oblique accretion stream incidence, the asymmetry is very evident and most of the plasma with temperature above 1 MK (thus the plasma which contributes to X-ray emission arising from the stream impact) is confined on the side of the stream in the direction of $v\rs{perp}$. 

\subsection{X-ray emission}

We have focused on the reference case (run D5e10-B30-A45), described in Fig.\ref{fig:model_dens} and Table \ref{tab1}, to derive detailed information regarding the X-ray emission of the shocked plasma from the accretion stream impacting onto the star, and regarding the effect of having an oblique stream onto such emission (with respect to the case of a stream normal to the star). To this aim, we have performed the synthesis of the X-ray emission that would result from the configuration considered in the reference case, applying for this the post-processing tool that we have developed and successfully used in the investigation of astrophysical shocks produced by streams (e.g. Bonito et al. 2011) and accretion streams (Bonito et al. 2014). The two cited references not only detail the tool that will be used here, but notably showed that the spectra synthesized from the numerical models can be directly compared with the astrophysical observations. 
We will here focus our synthesis of the radiation emanating from the modelled accretion impacts on the emission around the O VII and Ne IX triplets in X-rays, in particular exploring the effect of the local absorption on the emission at different wavelengths. The aim is to evaluate how the effects of this local absorption are changing when the structure of the accretion impact is distorted in the case of a slanted impact.
We have neglected the contribution to the emission due to the corona and the absorption due to the high temperature plasma and we use the Anders et al. abundances (as in Bonito et al. 2014). 
We find that the main contribution to the local absorption is the unperturbed accretion stream itself as well as the asymmetric lateral ejecta that perturb the ambient, as observed in the laboratory experiment and reproduced by the selected reference model here described. As the ejecta are asymmetric, what is find is that the absorption, and hence the emission, differ depending on the direction from which the emission would be collected from different sides of the accreting column.
We have synthesized the emission as emanating either from the left or from the right side of the accretion column, which is here modelled in two dimensions, in order to account for the two extremes of minimum and maximum contribution of the local absorption. Indeed, as is obvious from Fig.\ref{fig:model_dens}, the emission propagating to the right of the figure will encounter much less lateral ejecta than that propagating to the left, hence we expect that they will be differently affected by local absorption in the plasma surrounding the incoming accretion stream. 

\begin{figure}[htp]
    \centering
    \includegraphics[width=9 cm]{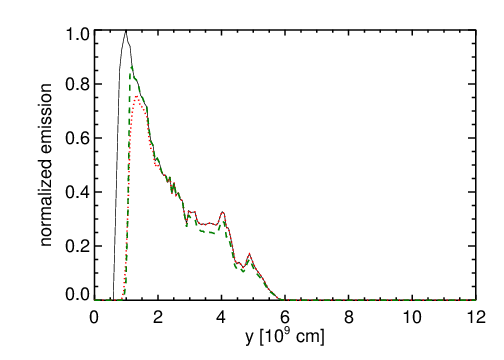}
    \caption{Simulated X-ray emission emanating from the accretion column modelled in run D5e10-B30-A45, as a function of the height above the star surface ($z$), for three cases: without any local absorption effect (black continuous line), and with the local absorption integrating the emission from the right (green dashed line) and from the left (red dotted line) side of the accretion stream.}
    \label{fig:image-1D}
\end{figure}

We have calculated the X-ray emission emanating from each point of the  computational domain, both neglecting local absorption and taking its effect into account. For the latter, we have weighted the emission by absorption induced by the plasma located between any point of the simulation grid and an observer that would be located either on the left or on the right side of the stream. 
Part of the emission (at the base of the shock) is totally absorbed due to the sinking of the shock into the chromosphere. 
Fig. \ref{fig:image-1D} shows the total emission emanating from the plasma, as a function of the height above the star surface. The emission is shown for three cases: without considering any absorption in the plasma (black continuous line), and considering absorption, but as seen either from the right (green dashed line) or the left (red doted line) of the column. The last two cases represent the minimum and maximum effect of the local absorption due to the accretion stream itself as well as due to the asymmetric lateral perturbations detected in both the laboratory experiment and the numerical model. We naturally observe in Fig. \ref{fig:image-1D} a reduction of the peak of the emission when considering absorption, with that reduction being different depending whether the accretion column is seen from the right or the left. Obviously, this is due to the asymmetric nature of the lateral ejecta. The peak corresponding to the bright blob located at the height $4\times 10^{9}$ cm is almost identical in the unabsorbed case and in the absorbed case synthesized from the left side of the accretion stream, (compare second peak in Fig. \ref{fig:image-1D} in black continuous line and red dotted line), while the main peak (the bright region at the base of the shock, but not sinked in the chromosphere) is strongly reduced due to the local absorption contribution of the lateral asymmetrical material. 
In the absorbed case synthesized from the right side of the accretion stream (green dashed line in Fig. \ref{fig:image-1D}), both the peaks of intensity (the one at the base, and the one located at $4\times 10^{9}$ cm) are strongly suppressed with respect to the unabsorbed case, but the main peak is higher with respect to the absorbed case observed from the left side of the shock (red dotted line in Fig. \ref{fig:image-1D}). In this case, the contribution to the emission due to the bright region at the base of the shock (not sinked into the chromosphere) is less absorbed than in the previous case as an effect of the asymmetric lateral structure formed as a consequence of the slanted accretion stream. The second peak of emission higher with respect to the stellar surface (at approximate $4\times 10^{9}$ cm, see Fig. \ref{fig:image-1D}) is more absorbed due to the local absorption of the accretion stream itself. The reduction of the main peak (located approximately at $1.5\times 10^{9}$ cm, see Fig. \ref{fig:image-1D}) with respect to the case without the local absorption taken into account corresponds to approx. $14\%$ in the absorbed case observed from the right side of the stream (green dashed line in Fig. \ref{fig:image-1D}) and to approx. $24\%$ in the absorbed case observed from the left side of the stream (red dotted line in Fig. \ref{fig:image-1D}), due to the asymmetrical structures formed in the slanted accretion stream. This result highlights the role of the local absorption and the importance of the lateral structures formed as a consequence of the inclination of the accretion stream with respect to the stellar surface. 
We have also simulated the spectra, in particular in correspondence of the relevant triplets for X-ray diagnostic (O VII and Ne IX). In both cases, we observe a reduction of the peak of the emission when the local absorption is taken into account, which is again different depending on the direction from which the observation is made. 

\section{Discussion and conclusions}
\label{sec:conc}
On the base of the experimental data it is shown:
(1)	Lack of confinement of plasma around the accretion column with increasing of the tilt of the incoming stream, confirmed by interferometry, SOP and VSG; (2) lateral escaping of the accreting plasma and increasing asymmetry of the accretion column, confirmed by interferometry and SOP (3); reduction of the temperature of the impacting plasma, confirmed by VSG.

All this is well supported by the astrophysical simulations presented above and can be summarized in terms of plasma-magnetic field interaction. Indeed, at impact, in the case of a stream impacting the obstacle normally, the laterally ejected plasma along the obstacle surface will encounter a perpendicular magnetic field, capable of stopping that lateral ejecta and refocus it at the edges of the incoming stream (\citealt{2017SciA....3E0982R}). However, when increasing the inclination of the obstacle, the ejected material will also encounter an increasingly oblique magnetic field, having a lesser capability of stopping the flow. All this leads to lesser plasma confinement, reduced heating, increased plasma propagation along the obstacle surface, in an increasingly asymmetric configuration.

Overall, both the experiments and the MHD simulations predict that the stellar
atmosphere in the immediate surrounding of an impact region of an
accretion stream can be heavily perturbed by the impact. According
to simulations and experiments, the perturbation increases with the
incidence angle especially on the side of the stream in the direction
of $v_{\rm perp}$ where a plasma motion parallel to the stellar
surface can be generated at the base of the stream. Depending on
the plasma $\beta$, the magnetic field can be significantly advected
by these outflows and the effect increases for higher incidence angles of the stream. In these cases, the conditions for ideal MHD may break down. Several authors have shown that perturbations traveling through the solar atmosphere and interacting with magnetic field lines can induce strong instabilities that lead to reconnective phenomena (e.g. \citealt{2007SoPh..246...89I, 2012ApJ...760L..10L, 2013ApJ...765...15J, 2015SSRv..190..103J}, and references therein). Thus, we expect that perturbations induced by the lateral flows may incite magnetic reconnection in proximity of the impact region, leading to a local release of the stored energy from the magnetic field that may heat up the surrounding stellar atmosphere (e.g. \citealt{2016ApJ...830...21R}). These effects are not described by our models as they do not include resistivity effects. From the experiments,
we do not see any indication of heating release, but we have to underline that the X-ray emission we record is dominated by the dense plasma on-axis, when possible heating would affect laterally ejected plasma having low-density, which would then little contribute to the overall emission. 

In addition to possible magnetic reconnection events
that may occur in proximity of impact regions, we expect that stream impacts with a significant incidence angle may easily trigger MHD waves which propagate in the chromosphere and which, again, may
perturb the structure of the chromosphere and, possibly, of the overlying corona. In the Sun MHD waves can be initiated as a consequence of several events as eruptive flares triggering oscillatory phenomena (e.g. \citealt{2004SoPh..223...77V, 2007ApJ...664.1210D, 2007A&A...473..959V, 2009A&A...508.1485V, 2013ApJ...777...17S, 2014ApJ...786..151S, 2015SSRv..190..103J}), or large-scale coronal streamers following the impact of a rapidly propagating coronal mass ejection (e.g.  \citealt{2010ApJ...714..644C, 2011ApJ...728..147C}). More recently, solar observations have recorded impacts of falling fragments after the eruption of a filament in a flare which have physical characteristics (infalling speed and density) close to those inferred for accretion flows on young accreting stars (\citealt{2013Sci...341..251R}). The analysis of observations and their comparison with hydrodynamic models describing these impacts have revealed a strong perturbation of the solar atmosphere (chromosphere and corona) and the triggering of oscillatory motion (\citealt{2014ApJ...797L...5R, 2016ApJ...832....2P, 2017A&A...598L...8P}). In our study, the generation of waves is expected to be the largest on the side of the stream in the direction of $v_{\rm perp}$, where the chromosphere is more perturbed by the stream impact.

The possibility that stream impacts may produce a significant perturbation 
of the stellar atmosphere has implications for the evidence that the 
observed coronal activity is apparently influenced by accretion (but it is
not clear why and how; \citealt{nss95, def09}). Some authors proposed that the coronal activity 
is modulated by mass accretion through the suppression, disruption,
or absorption of the coronal magnetic activity (e.g. \citealt{fdm03, sab04a, pkf05, jca06, gwj07}); others
proposed that the accretion may enhance the coronal activity around
impact regions due to heating of the surrounding stellar atmosphere to
soft X-ray emitting temperatures (e.g. \citealt{bcd10, dbc12}). Our analysis suggests
that accretion impacts with an incidence angle may produce significant
outflows at the base of accretion streams; we propose that these outflows may disrupt or suppress
coronal magnetic activity around impact regions. In case of magnetic
reconnection events in proximity of impact regions we expect that the coronal activity
may be enhanced. The perturbation of the stellar atmosphere may also
contribute to the stellar outflow as suggested by \cite{2008ApJ...689..316C} through a model
of accretion-driven winds in CTTSs (see, also, \citealt{2009ApJ...706..824C}). According
to this model, the mass loss rates observed in CTTSs can be explained if,
in addition to the convection-driven MHD turbulence which dominates in
solar-like stars, is present a source of wave energy driven by stream
impacts onto the stellar surface. Future modeling studies including
resistive effects and an accurate description of the density and
temperature structure of the stellar atmosphere (from the photosphere to
the chromosphere and to the corona) and of the stellar magnetic field
in proximity of impact regions may offer additional insight into the
reaction of the stellar atmosphere to accretion impacts and shed light on 
the connection between mass accretion rates and level of corona activity.

\begin{acknowledgements}
This work was supported by funding from the European Research Council (ERC) under the European Unions Horizon 2020 research and innovation program (Grant Agreement No. 787539), as well as ANR Blanc Grant n 12-BS09-025-01 SILAMPA (France) and by the Ministry of Education and Science of the Russian Federation under Contract No. 14.Z50.31.0007. This work was partly done within the LABEX Plas@Par, the DIM ACAV funded by the Region Ile-de-France, and supported by Grant No. 11-IDEX- 0004-02 from ANR (France). Part of the experimental system is covered by a patent (n 1000183285, 2013, INPI-France). The research leading to these results is supported by Extreme Light Infrastructure Nuclear Physics (ELI-NP) Phase II, a project co-financed by the Romanian Government and European Union through the European Regional Development Fund. The PLUTO code is developed at the Turin Astronomical Observatory (Italy) in collaboration with the Department of General Physics of Turin University (Italy) and the SCAI Department of CINECA (Italy). The astrophysical simulations were carried out at the SCAN (Sistema di Calcolo per l'Astrofisica Numerica) facility for high performance computing at INAF -- Osservatorio Astronomico di Palermo (Italy).
\end{acknowledgements}

\bibliographystyle{aa}
\bibliography{biblio}

\end{document}